\documentclass[fleqn,12pt]{wlscirep}
\usepackage[utf8]{inputenc}
\usepackage[T1]{fontenc}

\usepackage{cite}
\usepackage{fixltx2e}
\usepackage{datetime}
\usepackage{graphicx}
\usepackage{stfloats}
\usepackage{color}
\usepackage{authblk}
\usepackage{amssymb}
\usepackage{algorithm}
\usepackage{pstricks,epsfig}
\usepackage{rotating}
\usepackage{subfig}
\usepackage{tikz}
\usepackage{enumerate}
\usepackage{tabularx}
\usepackage{multirow}
\usepackage{verbatim}
\usepackage{pgf}
\usepackage{url}

\DeclareMathAlphabet{\mathsfsl}{OT1}{cmss}{m}{sl}

\title{Synthetic Correlated Diffusion Imaging Hyperintensity Delineates Clinically Significant Prostate Cancer}

\author[1,3,*]{Alexander Wong}
\author[1*]{Hayden Gunraj}
\author[1]{Vignesh Sivan}
\author[2,4]{Masoom A. Haider}
\affil[1]{Department of Systems Design Engineering, University of Waterloo, Waterloo, Canada}
\affil[2]{Joint Department of Medical Imaging,  Mount Sinai Hospital, Princess Margaret Hospital, University of Toronto, Toronto, Canada}
\affil[3]{Ontario Institute for Cancer Research, Toronto, Canada}
\affil[4]{Lunenfeld-Tanenbaum Research Institute, Ontario, Canada}
\affil[*]{\{a28wong,hayden.gunraj\}@uwaterloo.ca}

\begin{abstract}
Prostate cancer (PCa) is the second most common cancer in men worldwide and the most frequently diagnosed cancer among men in more developed countries.
The prognosis of PCa is excellent if detected at an early stage, making early screening crucial for detection and treatment.  In recent years, a new form of diffusion magnetic resonance imaging called correlated diffusion imaging (CDI) was introduced, and preliminary results show promise as a screening tool for PCa.  In the largest study of its kind, we investigate the relationship between PCa presence and a new variant of CDI we term synthetic correlated diffusion imaging (CDI$^s$), as well as its performance for PCa delineation compared to current standard MRI techniques (T2-weighted (T2w) imaging, diffusion-weighted imaging (DWI), and dynamic contrast-enhanced (DCE) imaging) across a cohort of 200 patient cases. Statistical analyses reveal that hyperintensity in CDI$^s$ is a strong indicator of PCa presence and achieves strong delineation of clinically significant cancerous tissue compared to T2w, DWI, and DCE.  These results suggest that CDI$^s$ hyperintensity may be a powerful biomarker for the presence of PCa, and may have a clinical impact as a diagnostic aid for improving PCa screening.
\end{abstract}

\begin{document}

\maketitle

\section*{Introduction}

Prostate cancer (PCa) is the second most common form of cancer in men worldwide and the most frequently diagnosed cancer among men in more developed countries, with roughly 1.4 million new cases in 2020~\cite{globocan21}.  With a median patient survival time for metastatic PCa ranges from 12.2 to 21.7 months~\cite{meta1,meta2,meta3,meta4,meta5}, early screening and diagnosis of PCa becomes critical for improving the treatment of patients with PCa, particularly since the five-year survival rate after diagnosis for patients with PCa at the non-metastatic stage is very high~\cite{survival}.  Clinical screening of PCa has traditionally involved the use of prostate-specific antigen (PSA) screening, with high PSA levels used as an indicator of PCa~\cite{Stenman}.  Unfortunately, studies have shown that PSA screening has led to a significant over-diagnosis of men suspected of PCa, resulting in over-treatment that carry significant risks~\cite{Chou,Loeb}.

While diagnostic imaging has been increasingly prevalent for PCa screening and diagnosis, one can argue that there is currently no universally accepted method for screening and diagnosing prostate cancer. Transrectal ultrasound (TRUS) is routinely used to guide prostate biopsy; however, its use for PCa screening and diagnosis is limited due to the fact that PCa tumours are often isoechoic and thus cannot be delineated from surrounding tissue via TRUS.  As a result, PCa screening and diagnosis using TRUS has low sensitivity and specificity in the range of 40-50\%~\cite{Norberg,Beerlage}. PCa screening and diagnosis using positron emission tomography (PET) has also been explored, with several tracers showing promise for delineating cancerous and non-cancerous tissue in the prostate gland~\cite{PET1,PET2,PET3,PET4}. Unfortunately, the high cost of PET scanning makes it impractical as diagnostic tool early in the screening pathway.

Magnetic resonance imaging (MRI) has grown significantly in prevalence for the purpose of PCa screening, with wide acceptance of the standardized PCa reporting system PI-RADS~\cite{PIRADS} centered around different MRI modalities. T2-weighted MRI (T2w) has been well-studied for PCa screening and diagnosis~\cite{T2a,T2b,T2c}, where potentially cancerous regions are characterized by signal hypointensity, and is considered the primary determining modality for the transition zone (TZ) in PI-RADS~\cite{PIRADS}. However, T2w signal hypointensity in the peripheral zone (PZ) of the prostate gland can also be associated with a number of non-cancerous abnormal conditions such as inflammation, fibrosis, and hemorrhage~\cite{fMRI}, leading to false positives if T2w was the sole method used.  To improve diagnostic accuracy when using MRI for PCa screening and diagnosis, two complementary MRI techniques have been leveraged for improved PCa screening alongside T2w: 1) diffusion-weighted imaging (DWI) with apparent diffusion coefficient (ADC) calculated from DWI, and 2) dynamic contrast-enhanced (DCE) imaging.  These techniques, when used together with T2w, form a multi-parametric MRI (MP-MRI) strategy to overcome the shortcomings of each modality. However, the need to interpret several modalities can increase interpretation challenges, resulting in increased inter- and intra-observer variability.

Recently, a new MRI technique called correlated diffusion imaging (CDI)~\cite{CDI} was proposed for improving PCa diagnosis. Preliminary studies demonstrated the potential of CDI for delineating between cancerous and non-cancerous tissue~\cite{CDI,Khalvati15}.  However, the scope of these studies are limited in terms of  patient cohort size and diversity (e.g., a patient cohort of 20 patient cases~\cite{CDI}). Furthermore, a number of limitations exist in CDI as first introduced with respect to signal-to-noise ratio (SNR) and acquisition time, as well as SI variability amongst inter-patient and intra-patient acquisitions.

The contribution of this study are two-folds.  First, this study represents the largest study of its kind for exploring the relationship between PCa presence and CDI signal hyperintensity across a cohort of 200 patient cases.  Second, an extended variant of CDI we term synthetic correlated diffusion imaging (CDI$^s$), which leverages a hybrid of native and synthetic diffusion signal acquisitions and signal calibration for greater consistency in dynamic range across machines and protocols, is introduced in this study, with its performance for PCa delineation compared to current standard MRI techniques (T2w imaging, DWI, and DCE imaging). This study aims to provide insights on the potential clinical impact of CDI$^s$ as a diagnostic aid for improving PCa screening.

\section*{Results}

In this study, we investigated the efficacy of CDI$^s$ from two different perspectives.  First, we studied the relationship between CDI$^s$ SI and the presence of PCa, both clinically significant PCa tissue and clinically insignificant PCa (Ins-PCa) tissue.  Consistent with the contemporary concept of clinically significant versus insignificant prostate cancer (PCa vs. Ins-PCa)~\cite{Ploussard11}, PCa tissue is defined as tissue with a Gleason score greater than or equal to 7 (Gleason Grade Groups 2-5 according to the International Society of Urological Pathology) while Ins-PCa tissue is defined as tissue with a Gleason score less than 7 (Gleason Grade Group 1).  Second, we studied the performance of CDI$^s$ in delineating PCa tissue and Ins-PCa tissue from healthy tissue.

~\\
\textbf{Relationship between CDI$^s$ SI and the presence of PCa.}
~\\
Fig. 1 shows the histogram analysis conducted to study the distribution of CDI$^s$ SI, T2w SI, DWI-derived ADC values, and DCE-derived $k^{trans}$ values for healthy tissue, PCa tissue, and Ins-PCa tissue.  A number of observations can be made from this histogram analysis. First, CDI$^s$ SI hyperintensity is clearly exhibited in the presence of PCa, with the clinical significance of PCa (from healthy tissue to PCa tissue) progressively increasing with the CDI$^s$ SI.  This observation means that not only can CDI$^s$ SI hyperintensity be a good indicator for the presence of PCa, but can also be a good risk assessment and treatment planning tool for quantitatively assessing the degree of disease severity.  Second, it can be observed that there is noticeably lower overlap between the CDI$^s$ SI distributions of PCa tissue and Ins-PCa tissue when compared to that of T2w SI, DWI-derived ADC values, and DCE-derived $k^{trans}$ values.  More specifically, the T2w SI distributions of healthy tissue, PCa tissue, and Ins-PCa tissue all have considerable overlap, while the value distributions of Ins-PCa tissue and PCa tissue have greater overlap for both $k^{trans}$ and ADC values.  This observation means that CDI$^s$ can potentially be a good clinical decision support tool for clinicians when compared to the current standard MRI techniques in determining the course of action for a patient, be it watchful waiting, active surveillance, or immediate treatment.

\begin{figure}
\centering\includegraphics[width=\textwidth]{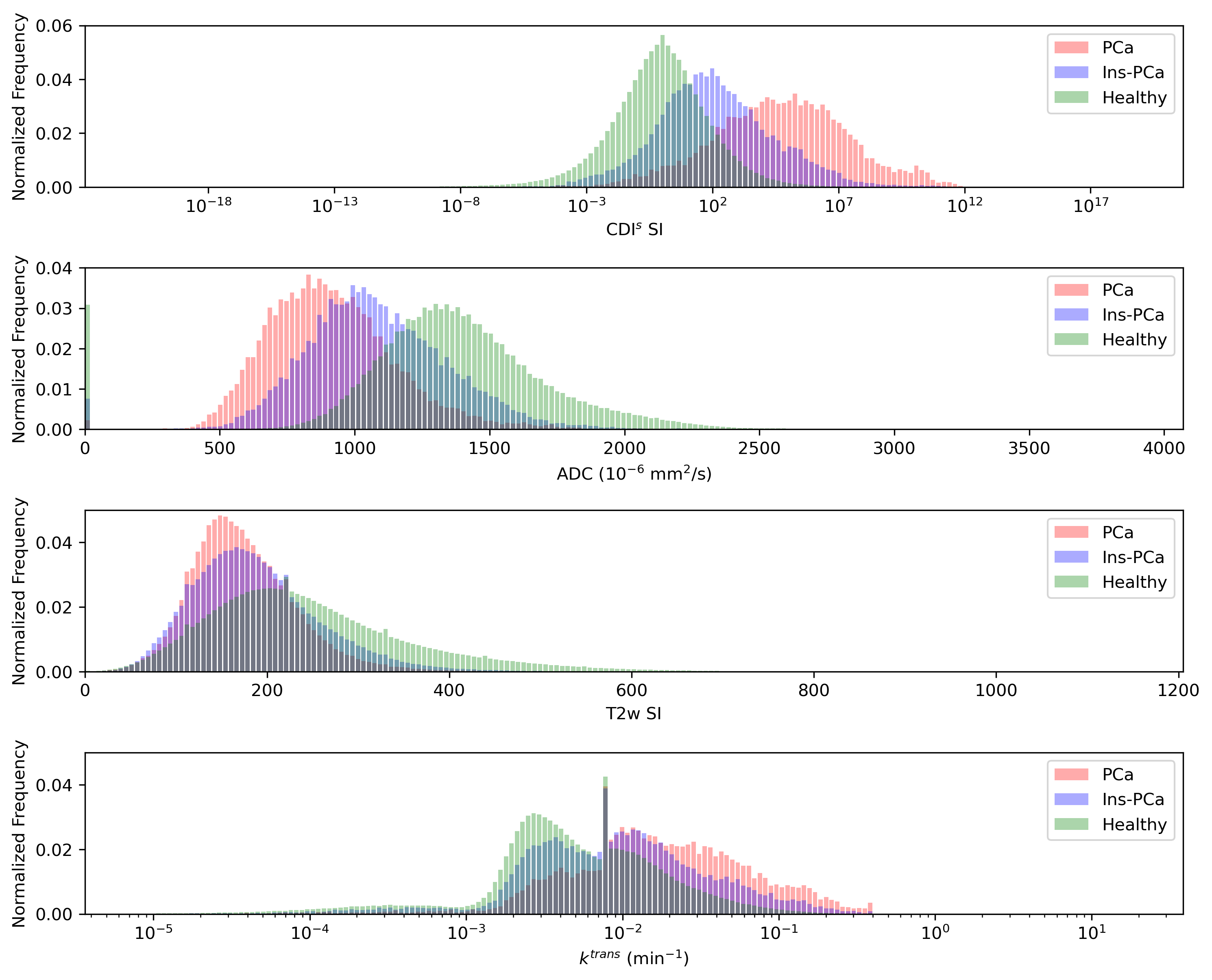}
\caption{Histogram analysis of CDI$^s$ SI, T2w SI, DWI-derived ADC values, and DCE-derived $k^{trans}$ values for healthy tissue, Ins-PCa, and PCa tissue.}
\label{fig:fig2}
\end{figure}

~\\
\textbf{Delineation between PCa tissue and healthy tissue based on quantitative analysis.}
~\\
Fig. 2 shows the ROC curves for studying the performance of CDI$^s$ SI, CDI$^s_t$ SI, T2w SI, DWI-derived ADC values, and DCE-derived $k^{trans}$ values for delineating PCa tissue and Ins-PCa tissue from healthy tissue. Differences in area under the curve (AUC) between modalities were assessed for statistical significance using the formulation proposed by Hanley and McNeil\cite{Hanley}, with the results of these tests given in Supplementary Table 1.

A number of observations can be made from this ROC analysis. First, it can be observed that  CDI$^s$ SI and CDI$^s_t$ SI achieve noticeably higher AUC for delineating between PCa tissue and healthy tissue when compared to the current standard MRI techniques, with DWI-derived ADC values achieving the next highest AUC (lower by $\sim$0.0308 when compared to  CDI$^s$ SI, $p=0.0000$). T2w and DCE-derived $k^{trans}$ values achieve significantly lower AUC compared to the other techniques (lower by as much as $\sim$0.2037 when compared to  CDI$^s$ SI, $p=0.0000$). When comparing  CDI$^s$ SI and  CDI$^s_t$ SI, it can be seen that CDI$^s_t$ SI achieves $\sim$0.0022 higher AUC ($p=0.0207$) when compared to CDI$^s$ SI, thus illustrating the efficacy of CDI$^s$ coefficient optimization where applicable.

Second, it can be observed that CDI$^s$ SI and CDI$^s_t$ SI achieve significantly greater delineation between PCa and Ins-PCa tissue when compared to current standard MRI techniques, with DWI-derived ADC values, T2w, and DCE-derived $k^{trans}$ values achieving $\sim$0.0612 ($p=0.0000$), $\sim$0.2017 ($p=0.0000$), and $\sim$0.1344 ($p=0.0000$) lower AUC when compared to CDI$^s$ SI, respectively. Here, the difference in delineation performance between PCa and Ins-PCa tissue for CDI$^s$ SI and CDI$^s_t$ SI is not significant ($p=0.8449$).

Third, it can be observed that CDI$^s$ SI and CDI$^s_t$ SI achieve noticeably higher AUC for delineating between PCa tissue and other tissue when compared to the current standard MRI techniques, with DWI-derived ADC values achieving the next highest AUC (lower by $\sim$0.0314 when compared to  CDI$^s$ SI, $p=0.0000$). T2w and DCE-derived $k^{trans}$ values achieve significantly lower AUC compared to the other techniques (lower by as much as $\sim$0.2032 when compared to  CDI$^s$ SI, $p=0.0000$). When comparing  CDI$^s$ SI and  CDI$^s_t$ SI, it can be seen that CDI$^s_t$ SI achieves $\sim$0.0021 higher AUC ($p=0.0251$) when compared to CDI$^s$ SI, thus again illustrating the efficacy of CDI$^s$ coefficient optimization where applicable.

Fourth, it can be observed that CDI$^s$ SI and CDI$^s_t$ SI achieve noticeably higher AUC for delineating between PCa tissue (both PCa and Ins-PCa) and healthy tissue when compared to the current standard MRI techniques, with DWI-derived ADC values achieving the next highest AUC (lower by $\sim$0.0151 when compared to CDI$^s$ SI, $p=0.0000$). T2w and DCE-derived $k^{trans}$ values achieve significantly lower AUC compared to the other techniques (lower by as much as $\sim$0.1510 when compared to  CDI$^s$ SI, $p=0.0000$). When comparing  CDI$^s$ SI and  CDI$^s_t$ SI, it can be seen that CDI$^s_t$ SI achieves $\sim$0.0063 higher AUC ($p=0.0000$) when compared to CDI$^s$ SI, thus again illustrating the efficacy of CDI$^s$ coefficient optimization where applicable.

\begin{figure}
\centering\includegraphics[width=\textwidth]{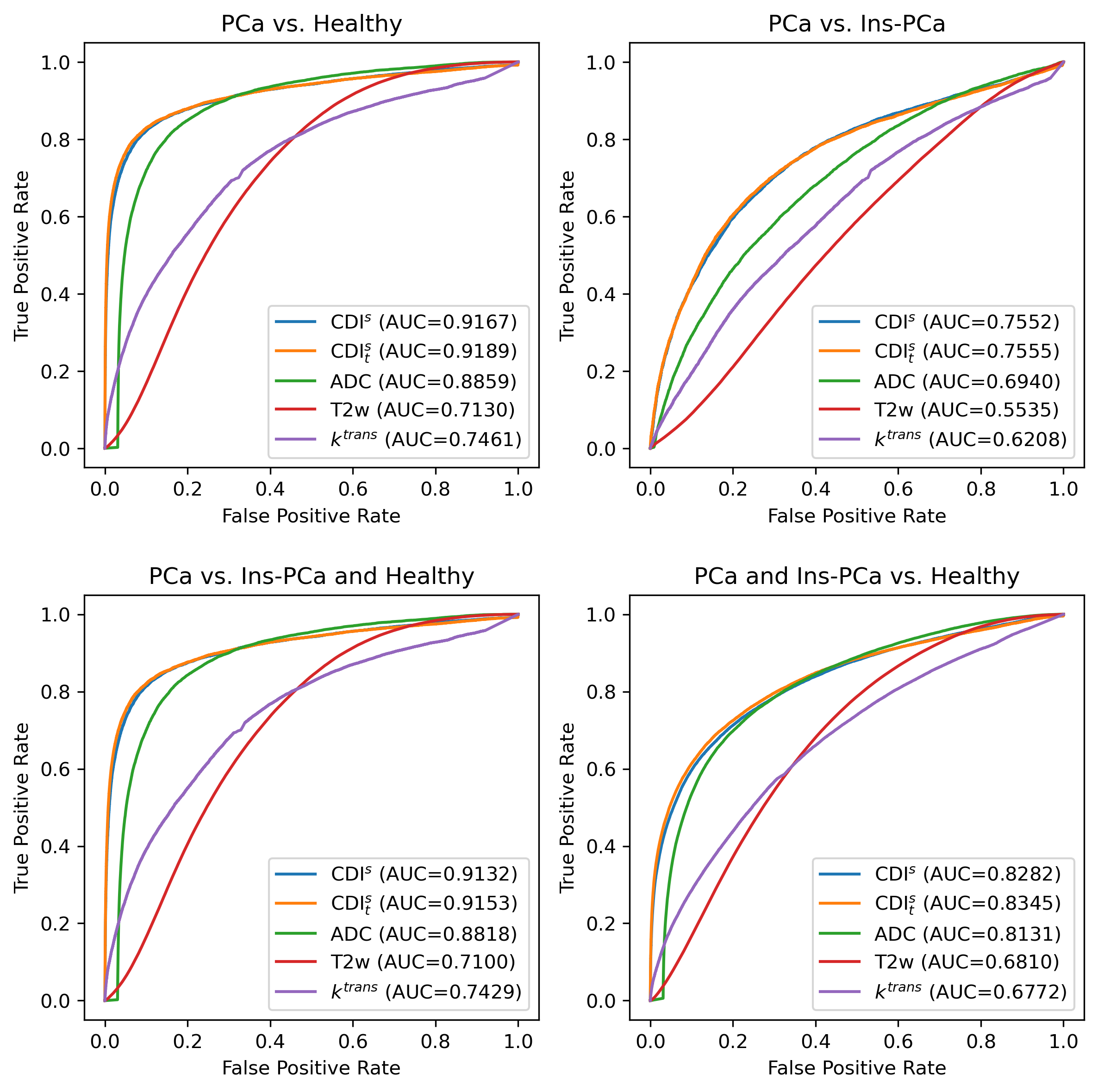}
\caption{ROC curves for studying the performance of CDI$^s$ SI, CDI$^s_t$ SI, T2w SI, DWI-derived ADC values, and DCE-derived $k^{trans}$ values for delineating PCa tissue and Ins-PCa tissue from healthy tissue.}
\label{fig:fig4}
\end{figure}

~\\
\textbf{Clinical interpretation.}
~\\
Fig. 3(a,b) shows the T2w and overlays of DWI-derived ADC, DCE-derived $k^{trans}$, and CDI$^s$ for two patient cases with PCa in the PZ.  In Fig. 3(a), it can be observed that T2w shows no contrast between PCa tissue and healthy tissue, while $k^{trans}$ exhibit strong contrast for a smaller portion within the PCa tumor.  ADC shows good contrast between the PCa tumor and some of the surrounding healthy tissue, but exhibit similar ADC values as the tumor in different small regions within the TZ, including an adjacent region above the tumor that makes it indistinguishable from the tumor itself.  CDI$^s$ shows strong contrast for the entire PCa tumor from the rest of the healthy tissue.

In Fig. 3(b), it can be observed that T2w shows poor contrast between PCa tissue and healthy tissue, while $k^{trans}$ exhibit poor contrast between the PCa tumor and healthy tissue.  ADC shows good contrast between the PCa tumor and surrounding healthy tissue but exhibit similar ADC values as the tumor in another small region within the PZ that was not identified as PCa tissue.  CDI$^s$ shows strong contrast for the entire PCa tumor from the rest of the healthy tissue.

Fig. 3(c,d) shows the T2w and overlays of DWI-derived ADC, DCE-derived $k^{trans}$, and CDI$^s$ for two patient cases with PCa in the TZ.  In Fig. 3(c), it can be observed that T2w shows no contrast between PCa tissue and healthy tissue.  $k^{trans}$ exhibit mild contrast in a small region within the PCa tumor, but strong contrast in a healthy tissue region that is not associated with PCa.  ADC shows strong contrast for the PCa tumor from surrounding tissue but exhibit similar ADC values as other regions within the PZ that were not identified as PCa tissue via histopathology validation.  CDI$^s$ shows strong contrast for the entire PCa tumor from the rest of the healthy tissue.  In Fig. 3(d), it can be observed that T2w shows no contrast between PCa tissue and healthy tissue, while $k^{trans}$, ADC, CDI$^s$ all exhibit strong contrast between PCa tumor and healthy tissue.

Fig. 3(e) shows the T2w and overlays of DWI-derived ADC, DCE-derived $k^{trans}$, and CDI$^s$ for a patient with PCa in the PZ and Ins-PCa in the TZ.  It can be observed that T2w shows no contrast between the PCa tumor and healthy tissue, and poor contrast between the Ins-PCa tumor and healthy tissue.  $k^{trans}$ exhibit no contrast between the PCa tumor and healthy tissue, and strong contrast for a small portion of the Ins-PCa tumor. ADC shows strong contrast between the PCa tumor and surrounding healthy tissue and good contrast between the Ins-PCa tumor and surrounding healthy tissue.  However, ADC exhibit similar values for both the clinically significant and clinically insignificant tumors, as well as exhibit similar ADC values as the tumor in small regions within the TZ that was not identified as PCa tissue.  CDI$^s$ show the strongest contrast between the PCa tumor and healthy tissue amongst the techniques, and shows good contrast between the Ins-PCa tumor and healthy tissue.  Furthermore,CDI$^s$ provides greater visual difference between PCa tumor and Ins-PCa tumor than ADC.

Fig. 3(f) shows the T2w and overlays of DWI-derived ADC, DCE-derived $k^{trans}$, and CDI$^s$ for a patient with  PCa in the anterior stroma (AS) and Ins-PCa in the PZ.  It can be observed that T2w shows poor contrast between the PCa tumor and healthy tissue, and T2w SI of the PCa tumor is similar to healthy tissue in the TZ.  T2s also provides good contrast between the Ins-PCa tumor.  T2w also shows good contrast between the Ins-PCa tumor and healthy tissue, although T2w SI of the Ins-PCa tumor is similar to healthy tissue in the TZ.  T2w also provides good contrast between the Ins-PCa tumor, but T2w SI in the Ins-PCa tumor is similar to that of healthy tissue in the TZ.  Furthermore, th T2w SI of the PCa tumor is very similar to that of the Ins-PCa tumor.  $k^{trans}$ exhibit no contrast between the PCa tumor and healthy tissue, and no contrast between the Ins-PCa tumor and surrounding healthy tissue.  Furthermore, $k^{trans}$ exhibit contrast in healthy tissue in the TZ that is not identified as PCa.  ADC shows poor contrast between the PCa tumor and surrounding healthy tissue and good contrast between the Ins-PCa tumor and surrounding healthy tissue.  CDI$^s$ show the strongest contrast between the PCa tumor and healthy tissue amongst the techniques, and shows good contrast between the Ins-PCa tumor and healthy tissue.  Furthermore, CDI$^s$ provides greater visual difference between PCa tumor and Ins-PCa tumor than ADC.

\begin{figure}
\centering\includegraphics[width=0.95\textwidth]{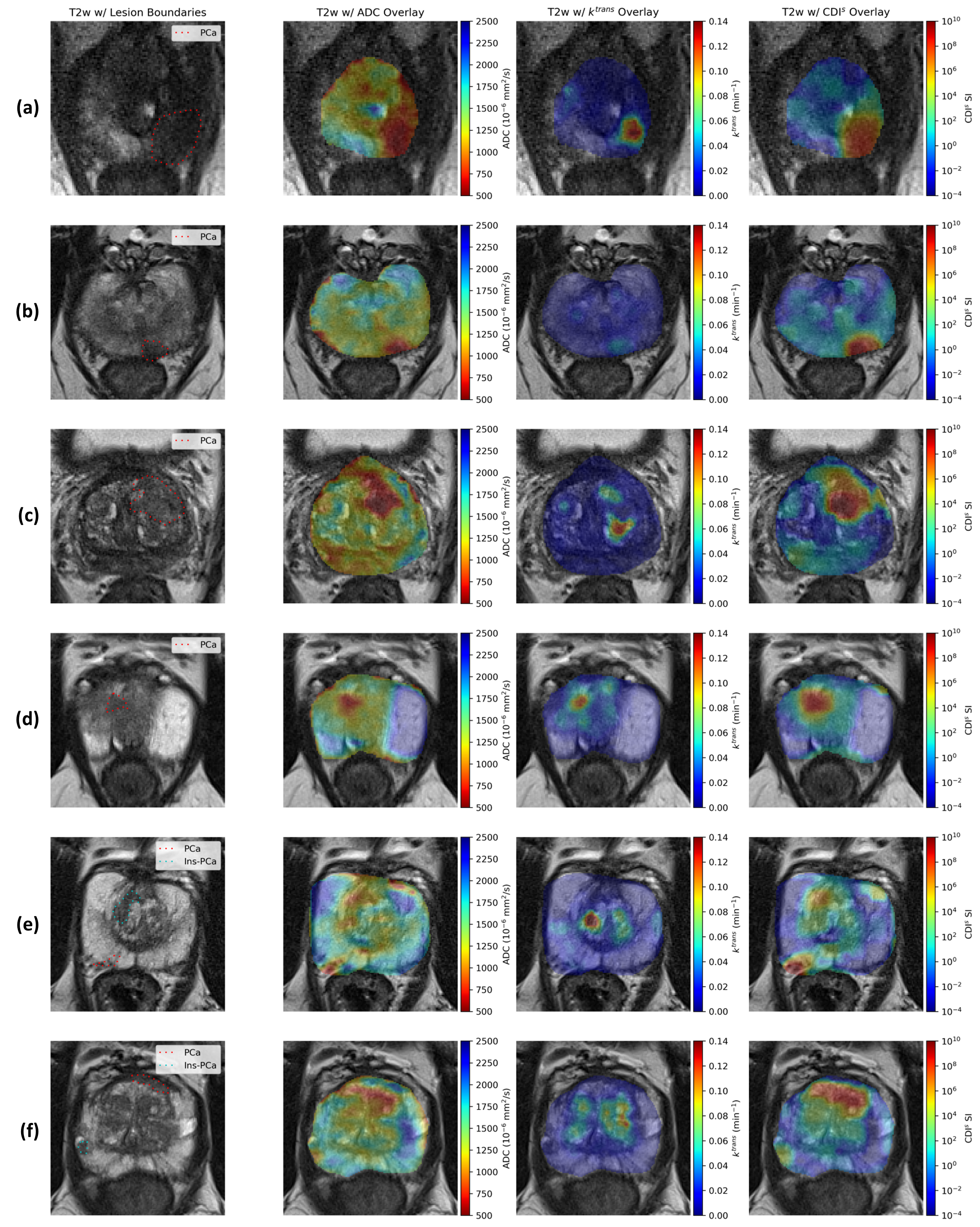}
\caption{T2w images with overlays of lesion boundaries, DWI-derived ADC, DCE-derived $k^{trans}$, and CDI$^s$ for six patient cases. \textbf{(a,b)} Two patients with PCa in the PZ. \textbf{(c, d)} Two patients with PCa in the TZ. \textbf{(e)} A patient with PCa in the PZ and Ins-PCa in the TZ. \textbf{(f)} A patient with PCa in the AS and Ins-PCa in the PZ.}
\label{fig:fig5}
\end{figure}

~\\
\section*{Discussion}

In this study, we hypothesised that there is a strong relationship between SI of CDI$^s$ and the presence of PCa, and the experimental results support this hypothesis. Results across a cohort of 200 patient cases with histopathology validation showed that hyperintensity in CDI$^s$ provides a strong indicator of the presence of clinically significant PCa.  Furthermore, CDI$^s$ achieves strong delineation of clinically significant cancerous tissue and healthy tissue (AUC exceeding 0.918 and 0.916 for CDI$^s_t$ and CDI$^s$, respectively), which is noticeably higher than current standard techniques for prostate screening such as T2w, DWI, and DCE. In general, CDI$^s$ also shows fewer false positive regions compared to the other techniques. These results suggest that the use of CDI$^s$ may have a clinical impact as a diagnostic aid for improving PCa screening.

To improve diagnostic accuracy when using MRI for PCa screening and diagnosis, DWI is often used alongside T2w, with DWI being the primary determining modality for the PZ in PI-RADS~\cite{PIRADS}.  In DWI, pairs of opposing magnetic field gradient pulses are applied in the imaging sequence to obtain sensitivity to the Brownian motion of water molecules in tissues~\cite{Koh}.  Therefore, given the presumed higher cellular density of cancerous tissue compared to non-cancerous tissue, potentially cancerous regions would exhibit signal hypointensity on ADC map~\cite{PIRADS} produced using DWI due to restricted diffusion.  Despite its considerable promise~\cite{Hosseinzadeh,Woodfield,Turkbey,Haider,T2d}, the use of DWI for PCa screening and diagnosis remains a challenge due to considerable ADC variability depending on the strength, duration, and timing of applied diffusion gradient pulses used in the DWI pulse sequences, thus necessitating fine-tuning the strength, duration, and timing of the applied diffusion gradient pulses. This is further complicated by significant overlap in ADC between stromal benign prostatic hyperplasia (BPH), anterior fibromuscular stroma (AFMS), central zone (CZ), and PCa~\cite{Quon}.

Another well-established modality leveraged to improve diagnostic accuracy when using MRI for PCa screening and diagnosis is DCE imaging~\cite{DCE}.  Here, a contrast agent in the form of low molecular-weight gadolinium chelate in injected intravenously, and T1-weighted MRI (T1w) is acquired before, during, and after the injection.  Given the increased permeability of the tumor vessels, potentially cancerous regions would exhibit a high volume transfer constant between blood plasma and the extravascular extra-cellular space (denoted by $k^{trans}$).  The use of DCE for PCa screening and diagnosis remains a challenge due to sensitivity to patient motion, is relatively non-specific~\cite{DCE}, and introduces additional acquisition complexities such as cost and process overhead due to the use of an agent.  As such, DWI, T2w, and DCE are frequently used together in the form of MP-MRI to overcome the shortcomings of each modality; however, the need to interpret multiple modalities also increases the difficulty in interpretation, leading to increased inter- and intra-observer variability.

To address the aforementioned shortcomings of DWI and ADC maps for PCa screening and diagnosis, a new diffusion MRI modality was recently introduced in the form of CDI~\cite{CDI}. In CDI, a series of pulse sequences with different gradient pulse strengths and timings are used to probe water molecules with different degree of Brownian motion in the tissues within a local volume.  Signal mixing is then performed on the signal acquisitions captured using these pulse sequences to determine the joint correlation of the acquisitions within a local volume.  As such, CDI leverages the distribution of water molecules with different degrees of Brownian motion in the tissues within the local volume to delineate between cancerous tissue (indicated by signal hyperintensity due to a wider spread in the distribution of water molecules with varying degrees of Brownian motion within a local volume) and non-cancerous tissue (indicated by lower relative intensity due to a tighter distribution of water molecules with a similar degree of Brownian motion within a local volume).  While preliminary studies have shown that CDI holds considerable promise of achieving greater signal delineation between cancerous and non-cancerous tissue when used as a standalone diagnostic imaging method~\cite{CDI} and when used in combination with T2w and DWI~\cite{Khalvati15}, these studies are rather limited in scope as the patient study sizes and patient diversity was relatively small.  Furthermore, CDI as it was originally investigated has several limitations associated with acquisition time and SNR-associated restrictions, and variability in SI amongst inter-patient and intra-patient acquisitions.  As such, a comprehensive study with a significantly larger patient size as well as extensions to CDI to address the aforementioned limitations is highly desired to achieve a thorough investigation and evaluation on the relationship between signal hyperintensity in CDI and presence of PCa, which was the basis of this study.

In conclusion, our results in this study support the hypothesis that the use of CDI$^s$ can be an effective tool for PCa screening and diagnosis, although additional studies are needed before adoption for routine clinical use. Furthermore, given the promising results, we aim to investigate the relationship of CDI$^s$ SI and the presence of other forms of cancer such as breast cancer, gastric cancer, and glioblastoma.

\section*{Methods}
~\\
\textbf{Imaging Protocol}.
~\\
To study the relationship between CDI$^s$ SI and PCa, a cohort of 200 patient cases with histopathology validation acquired at the Radboud University Medical Centre (Radboudumc) in the Prostate MR Reference Center in Nijmegen, The Netherlands~\cite{Litjens14} were used in this study.  Table 1 summarizes the demographic, MR scanner, and clinical significance variables of the patient cohort used in this study.  The patients in this cohort ranged in age from 37-78~years, with a median age of 64~years.  All acquisitions were reviewed by or under the supervision of an expert radiologist with over 20 years of experience  interpreting prostate MRI~\cite{Litjens14}.

All examinations were performed using a Siemens MAGNETOM Trio 3.0T machine or a Siemens MAGNETOM Skyra 3.0T machine at the Radboud University Medical Centre, Nijmegen, The Netherlands.  A single-shot echo-planar sequence was used for DWI acquisitions with the following imaging parameters: TR range from 2500-3300~ms with a median of 2700~ms, and TE ranged from 63-81~ms with a median of 63~ms. The resolution of the acquisitions was 2~mm in-plane and slice thickness ranged from 3-4.5~mm with a median of 3~mm, and Q=$\{50~\text{s/mm}^2, 400~\text{s/mm}^2, 800~\text{s/mm}^2\}$.  The display field of view (DFOV) ranged from 16.8$\times$25.6~cm$^2$ to 24.0$\times$25.6~cm$^2$ with a median of 16.8$\times$25.6~cm$^2$.

To compare the performance of CDI$^s$ for PCa delineation with current standard MRI techniques, ADC maps were also obtained from DWI acquisitions.  Furthermore, axial T2w acquisitions were obtained as a reference of comparison as well.  Examinations using T2w were performed using a turbo spin-echo sequence with the following imaging parameters: TR range from 3880-7434.8~ms with a median of 5660~ms, and TE ranged from 101-112~ms with a median of 104~ms. The resolution of the acquisitions ranged from 0.3-0.6~mm in-plane with a median of 0.5~mm and slice thickness ranged from 3-4.5~mm with a median of 3~mm.  The DFOV ranged from 18$\times$18~cm$^2$ to 19.2$\times$19.2~cm$^2$ with a median of 19.2$\times$19.2~cm$^2$.  Finally, DCE imaging was conducted with a turbo flash gradient-echo sequence  with the following imaging parameters: TR range from 3.72-36~ms with a median of 36~ms, and TE ranged from 1.41-1.84~ms with a median of 1.41~ms. The resolution of the acquisitions ranged from 1.3-1.8~mm in-plane with a median of 1.5~mm and slice thickness ranged from 3-5~mm with a median of 3.5~mm. The DFOV ranged from 19.2$\times$19.2~cm$^2$ to 25$\times$25~cm$^2$ with a median of 19.2$\times$19.2~cm$^2$. Maps of the pharmacokinetic parameter $k^{trans}$ were obtained from the DCE series.

PCa, whole gland, transition, and PZ annotations for all patient acquisitions in this cohort were used in this study, with the annotation being performed by two radiology residents and two experienced board-certified radiologists~\cite{Cuocolo21}. Clinical interpretation of CDI$^s$, T2w, ADC, and $k^{trans}$ was conducted in this study by an expert radiologist with over 20 years of experience interpreting prostate MRI (MAH).

\begin{table}[h]
\centering
\caption{Summary of demographic, MR scanner, and clinical significance variables of the patient cohort used in this study. Age and MR scanner statistics are expressed on a patient level, while clinical significance statistics are expressed on a tumor level.}
\label{tab:demographic}
\begin{tabular}{|c|c|}
\hline
\multicolumn{2}{|l|}{\textbf{Age}}\\ \hline
30-39 & 3 (1.5\%) \\ \hline
40-49 & 5 (2.5\%) \\ \hline
50-59 & 45 (22.5\%) \\ \hline
60-69 & 112 (56\%) \\ \hline
70-79 & 35 (17.5\%) \\ \hline
\multicolumn{2}{|l|}{\textbf{MR Scanner (Siemens MAGNETOM)}} \\ \hline
Skyra 3.0T & 195 (97.5\%) \\ \hline
Trio 3.0T & 5 (2.5\%) \\ \hline
\multicolumn{2}{|l|}{\textbf{Clinical significance (Gleason Score)}} \\ \hline
PCa (GS $\geq$ 7) & 76 (25.4\%) \\ \hline
Ins-PCa (GS < 7 or PI-RADS~\cite{PIRADS} = 2) & 223 (74.6\%) \\ \hline
\end{tabular}
\end{table}

~\\
~\\
\textbf{Synthetic Correlated Diffusion Imaging}.
~\\
In this study, an extended variant of CDI we term CDI$^s$ is introduced.  The first key distinguishing aspect of CDI$^s$ when compared to CDI is the introduction of synthetic signal acquisitions alongside native signal acquisitions to reduce acquisition time, allow existing clinical imaging protocols and pulse sequences that are routine in MP-MRI imaging sessions to be used, as well as overcome SNR limitations and distortion limitations faced by CDI, particularly under gradient pulse configurations with longer echo times.  The second key distinguishing aspect of CDI$^s$ when compared to CDI is the introduction of signal calibration into the signal mixing procedure of CDI to allow for greater consistency in the resulting SI dynamic range across machines and protocols.  This signal calibration thus addresses the issue associated with CDI with respect to large variability in SI amongst inter-patient and intra-patient acquisitions that could highly affect clinical interpretation.

\begin{figure}
\centering\includegraphics[scale=0.5]{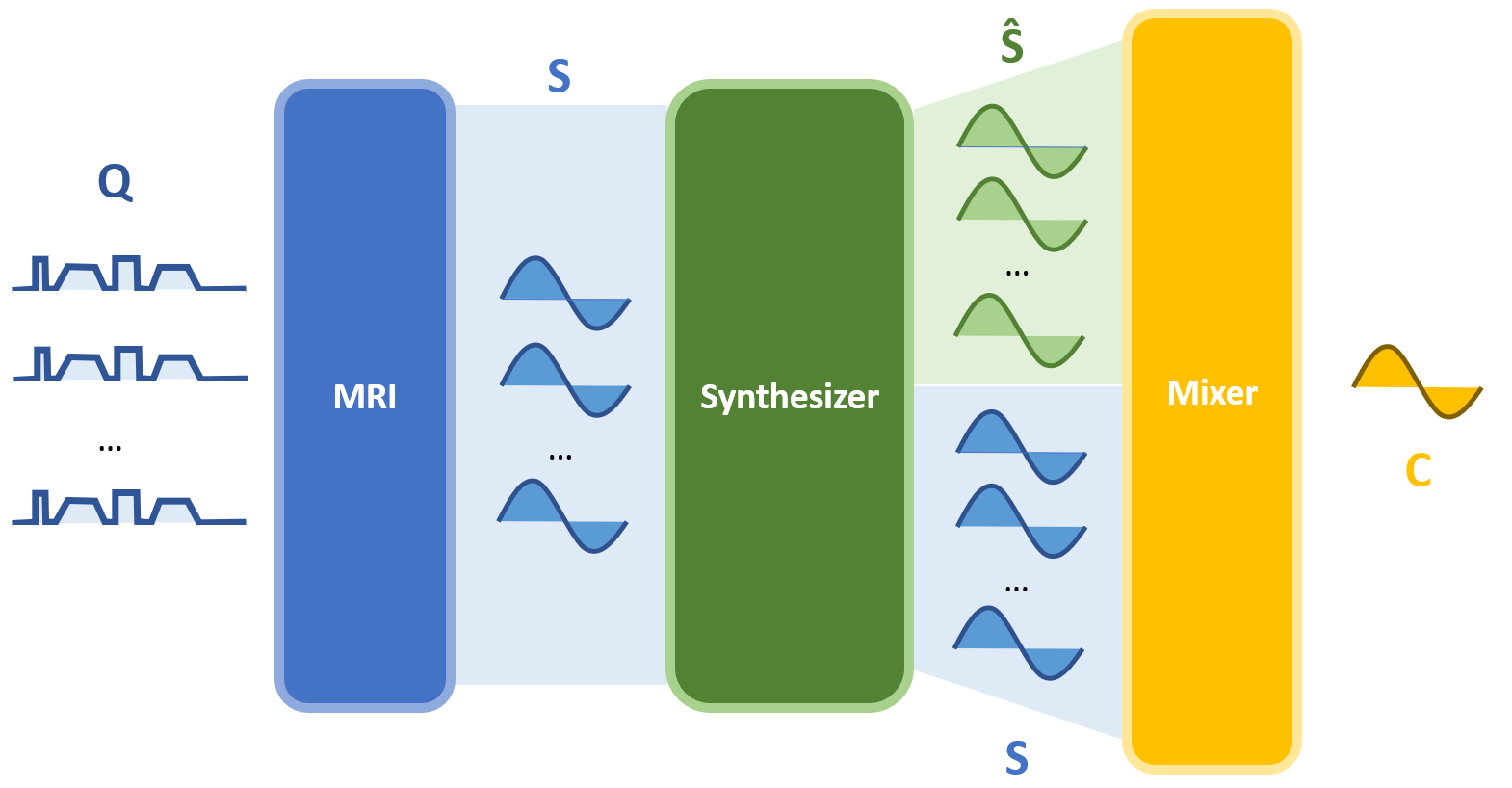}
\caption{The methodology behind synthetic correlated diffusion imaging. The methodology behind CDI$^s$ can be summarized as follows. First, multiple signal acquisitions are conducted using sequences with different gradient pulse strengths and timings $(q_1, q_2, …, q_N)$. Second, the native signal acquisitions $S$ are passed into a signal synthesizer to synthesize synthetic signal acquisitions $\hat{S}$.  Third, the native and synthetic signal acquisitions are then mixed together in a calibrated manner to obtain the local correlation of signal attenuation across the acquired signals, which produces a final signal (C) that characterizes the tissue being imaged with greater consistency in dynamic range across machines and protocols.}
\label{fig:scdi}
\end{figure}

The methodology behind CDI$^s$ is summarized in Fig. 4. First, multiple signal acquisitions using the aforementioned single-shot echo-planar sequence are conducted using a set of different configurations of gradient pulse strengths and timings, which we will denote as $Q=\{ q_i | i=1,...,N \}$, where $q_i$ denotes the $i^{\rm th}$ sequence.  By varying the configuration of gradient pulse strengths and timings between signal acquisitions, each signal acquisition is sensitive to a different degree of Brownian motion of water molecules in tissues within a local volume, thus allow the multiple signal acquisitions to provide a more complete, finer grain picture on the tissue characteristics within a local volume by quantifying the distribution of water molecules with respect to their degree of Brownian motion within tissue.  The different configurations of gradient pulse strengths and timings used in CDI$^s$ can be defined by~\cite{Stejskal},

\begin{equation}
q_i=(G_i, \delta_i, \Delta_i),
\end{equation}

\noindent where, for the $i^{\rm th}$ sequence, $G_i$ represents the gradient pulse strength, $\delta_i$ represents the gradient pulse duration, and $\Delta_i$ represents time between gradient pulses.  The configuration of gradient pulse strengths and timings used for a particular sequence $q_i$ can be simplified by grouping these terms~\cite{Bihan}, resulting in

\begin{equation}
q_i=\gamma^2 G_i^2\delta_i^2 \big( \Delta_i-\frac{\delta_i}{3}\big),
\end{equation}

\noindent where $\gamma$ denotes the proton gyromagnetic ratio.

Second, the multiple acquired signals are then passed into a signal synthesizer to synthesize synthetic signal acquisitions at desired configurations of gradient pulse strengths and timings not captured via native signal acquisitions.  Third, the native and synthetic signal acquisitions are mixed together in a calibrated manner to obtain the final quantitative signal characterizing the joint correlation \textbf{across} the acquired signals within a local volume $V$.  The idea behind this calibrated signal mixing procedure in CDI$^s$ stems from our hypothesis that the distribution of water molecules with different degrees of Brownian motion within tissue in a local volume with PCa would differ significantly from that with non-cancerous tissue.  For example, healthy PZ regions are largely comprised of glandular tissue, resulting in tighter distributions characterized by high relative SI at gradient pulse configurations with lower $q$ values and low relative SI at gradient pulse configurations with higher $q$ values used in CDI$^s$.  More importantly, BPH, AFMS, and CZ are all non-cancerous and are largely comprised of non-glandular tissue, resulting in distributions characterized by low relative SI across all gradient pulse strengths and timings used in CDI$^s$.  On the other hand, PCa regions are characterized by a more heterogeneous mixture of tissue with various degrees of relatively high cellular densities, resulting in a wider spread in the distribution of water molecules with varying degrees of Brownian motion within a local volume, including slow moving water molecules due to various degrees of true restricted diffusion, and thus high relative SI across all gradient pulse strengths and timings used in CDI$^s$.  In the case of DWI, BPH, AFMS, CZ, and PCa would exhibit ADC hypointensity, resulting in significant ADC overlaps between them, particularly depending on the choice of DWI sequence used, and thus increased risk of false positives~\cite{Quon}.  By taking advantage of these distribution differences between PCa and non-cancerous tissue in the signal mixing process in the form of joint correlation in a calibrated fashion, CDI$^s$ can achieve improved signal contrast between PCa and non-cancerous tissue and facilitate for more effective PCa screening and diagnosis, as well as greater consistency in SI dynamic range across machines and protocols to reduce inter-patient and intra-patient variability in clinical interpretation.

As mentioned earlier, we extend upon the signal mixing process in CDI~\cite{CDI} in two key ways to form CDI$^s$.  First, we introduce a calibrated signal mixing function $C(\underline x)$ for characterizing the contribution-adjusted local signal correlation across the multiple signal acquisitions, which is parameterized by a set of gradient pulse configurations $\{q_{\alpha},...,q_{\beta}\}$ and is defined as

\begin{equation}
C_{\{q_{\alpha},...,q_{\beta}\}}(\underline x)=\frac{1}{Z}\int \dots \int {S_{q_{\alpha}}(\underline x')^{\rho_\alpha}\dots S_{q_{\beta}}(\underline x')^{\rho_\beta}}f\left(S_{q_{\alpha}}(\underline x'),\dots,S_{q_{\beta}}(\underline x')|V(\underline x)\right)dS_{q_{\alpha}}(\underline x') \dots dS_{q_{\beta}}(\underline x'),
\end{equation}

 \noindent where $\underline x$ represents spatial location, $S$ represents a signal acquisition, $f$ presents the conditional joint probability density function, $V(\underline x)$ represents a local volume around $\underline x$, $\rho_i$ represents coefficients for controlling the contribution of the different gradient pulse strengths and timings, and $Z$ represents a calibration factor.  The calibration factor allow the signal mixing function to compensate for inherent variations due to the scanner machine and imaging protocol  used during acquisition which can lead to differences in CDI SI appearance across patients or even for the same patient at different acquisitions.

 For this study, $\{q_{\alpha},...,q_{\beta}\}$ was set at $\{50~\text{s/mm}^2,1000~\text{s/mm}^2,...,7000~\text{s/mm}^2\}$ (at $1000~\text{s/mm}^2$ intervals), $V$ was defined as a 6~mm~$\times$~6~mm~$\times$~3~mm volume centered at $\underline x$, and the definition of $\rho$ will be discussed in a later section.  The probability density function $f$ was defined as an uncorrelated Gaussian distribution with mean $\underline x$ and covariance matrix $\Sigma=\textbf{diag}(4~\text{mm}^2,4~\text{mm}^2,0~\text{mm}^2)$, and the calibration factor $Z$ is computed as the median CDI$^s$ SI within the prostate gland. These definitions yield the specific form of Eq.~3 used in this study:

\begin{equation}
C_{\{q_{\alpha},...,q_{\beta}\}}(\underline x)=\frac{1}{Z}\underset{V(\underline x)}{\iiint} {S_{q_{\alpha}}(\underline x')^{\rho_\alpha}\dots S_{q_{\beta}}(\underline x')^{\rho_\beta}}f\left(\underline x';\underline x, \Sigma\right)d\underline x'
\end{equation}

Second, we introduce the hybrid use of both native and synthetic signal acquisitions in the signal mixing process, thus leading to the notion of CDI$^s$.  The use of synthetic signal acquisitions alongside native signal acquisitions enables us to reduce the acquisition time required to capture signals at gradient pulse configurations with higher $q$ values, as well as overcome SNR limitations with acquiring signals at gradient pulse configurations with higher $q$ values due to factors such as longer echo times and eddy current-induced distortions.  Furthermore, it enables the leveraging of existing clinical imaging protocols and pulse sequences that are routine in MP-MRI imaging sessions.  More specifically, a synthetic signal acquisition $\hat S_q$ at gradient pulse configuration with a particular $q$ value can be synthesized at the signal synthesizer as

\begin{equation}
\hat S_q(\underline x) = S_{q_{ref}}(\underline x) \exp\left(-(q-q_{ref})\hat{A}(\underline x)\right),
\end{equation}
where $\hat{A}(\underline x)$ is the least-squares ADC estimate given by Eq.~6:

\begin{equation}
    \hat{A}(\underline x) = -\frac{\sum\limits_{q_i\in Q}(q_i-q_{ref})\log\left(\frac{S_{q_i}(\underline x)}{S_{q_{ref}}(\underline x)}\right)}{\sum\limits_{q_i\in Q}(q_i-q_{ref})^2}
\end{equation}
\noindent In this study, native signal acquisitions $S$ at $Q=\{50~\text{s/mm}^2,400~\text{s/mm}^2,800~\text{s/mm}^2\}$ were leveraged by the signal synthesizer to synthesize the aforementioned synthetic signal acquisitions $\hat{S}$ at $\{1000~\text{s/mm}^2,...,7000~\text{s/mm}^2\}$ (at $1000~\text{s/mm}^2$ intervals).  The synthesized signal acquisitions are then used alongside the native signal acquisition at $q=50~\text{s/mm}^2$ in the signal mixing process for CDI$^s$.
~\\
~\\
\textbf{CDI$^s$ Coefficient Optimization.}
~\\
As described in Eq. 3, the contribution of different gradient pulse strengths and timings to the CDI$^s$ signal produced by the signal mixer can be controlled via coefficients $\rho$.  In this study, we will study the efficacy of CDI$^s$ in both a baseline form (i.e., $\rho=1$ for all gradient pulse strengths and timings) as well as in a form that is tuned specifically to optimize delineation. More specifically, we tune the coefficients $\rho$ for the different gradient pulse strengths and timings using a Nelder-Mead Simplex optimization strategy, with the objective function being the area under the ROC curve. This coefficient optimization will yield a tuned form of CDI$^s$, which we will denote CDI$^s_t$, that accounts for the importance of the different gradient pulse strengths and timings to delineation performance.
~\\
~\\
\textbf{Visualization of CDI$^s$.}
~\\
To better map CDI$^s$ SI in a form that is more natural for clinical interpretation, the CDI$^s$ SI is transformed to the logarithmic space for clinical visualization purposes.  Given that CDI$^s$ is computed as a product of exponential signal acquisitions, CDI$^s$ SI is more naturally interpreted in a logarithmic space.  Furthermore, the transformed CDI$^s$ SI is visualized as a heatmap overlay on T2w images to provide additional anatomical context with respect to the prostate gland.  All image visualizations of CDI$^s$ SI in this study are shown with the aforementioned transforms.  Finally, DWI-derived ADC and DCE-derived $k^{trans}$ are visualized as heatmap overlays on T2w images for comparison purposes in this study.
~\\
~\\
\textbf{Statistical Analysis}.
~\\
Two different analysis methods were utilized in this study to investigate the relationship between CDI$^s$ SI and the presence of PCa, as well as the performance of CDI$^s$ for delineating between PCa tissue, Ins-PCa tissue, and healthy tissue.  In the first analysis method, we study the relationship between CDI$^s$ SI and the presence of PCa by performing histogram analysis to study the distribution of CDI$^s$ SI for healthy tissue, PCa tissue, and Ins-PCa tissue.  In the second analysis method, a receiver operating characteristic (ROC) curve analysis was performed using CDI$^s$ to quantitatively assess the ability to delineate between healthy tissue, clinical significant PCa tissue, and Ins-PCa tissue.  Consistent with the contemporary concept of significant versus insignificant prostate cancer (PCa vs. Ins-PCa)~\cite{Ploussard11}, Consistent with the contemporary concept of significant versus insignificant prostate cancer (PCa vs. Ins-PCa)~\cite{Ploussard11}, PCa tissue is defined as tissue with a Gleason score greater than or equal to 7 (Gleason Grade Groups 2-5 according to the International Society of Urological Pathology) while Ins-PCa tissue is defined as tissue with a Gleason score less than 7 (Gleason Grade Group 1).  The ROC curves were estimated assuming bivariate normal data.  For illustrative purposes, the ROC curves obtained from the pooled data of all patient cases was plotted.  To provide a quantitative assessment of diagnostic accuracy, the area under the ROC curve was obtained as a single metric of delineation performance. For comparison purposes, histogram analysis and ROC curve analysis were also performed using T2w, DWI-derived ADC map values, and DCE-derived $k^{trans}$ map values.

To assess the statistical significance of differences in AUC values between different modalities, we adopt the critical ratio formulation of Hanley and McNeil\cite{Hanley}. Specifically, for each pair of modalities, the critical ratio $z$ is defined as:

\begin{equation}
    z = \frac{A_1-A_2}{\sqrt{SE_1^2 + SE_2^2 - 2rSE_1 SE_2}},
\end{equation}

where $A_1$ and $SE_1$ denote to the AUC and estimated standard error (SE) of modality 1, $A_2$ and $SE_2$ denote the AUC and estimated SE of modality 2, and $r$ denotes the estimated correlation between $A_1$ and $A_2$. The estimated SE for each modality is defined as:
\begin{equation}
    SE_i = \sqrt{\frac{A_i(1-A_i) + (n_P - 1)\left(\frac{A_i}{2 - A_i} - A_i^2\right) + (n_N - 1)\left(\frac{2A_i^2}{1 + A_i} - A_i^2\right)}{n_Pn_N}},
\end{equation}
where $n_P$ and $n_N$ denote the number of true positive and true negative examples, respectively.

To estimate the correlation coefficient $r$, the Pearson product-moment correlation is first used to estimate correlation between the two modalities for positive examples ($r_P$) and negative examples ($r_N$). The average of $r_P$ and $r_N$ and the average of $A_1$ and $A_2$ are then used to linearly interpolate the standard table presented by Hanley and McNeil\cite{Hanley} to estimate $r$. Notably, differences in image resolution prevent one-to-one comparison of voxels when estimating the Pearson product-moment correlation, and as such the median values for PCa, Ins-PCa, and healthy tissue are used (computed per-patient).
~\\
~\\
\textbf{Ethics}.
~\\
This study has received ethics clearance from the University of Waterloo (30632), and was carried out in accordance with relevant guidelines and regulations. Informed consent was obtained from all subjects and/or their legal guardian(s).




\section*{Acknowledgments}

The authors thank the Natural Sciences and Engineering Research Council of Canada and the Canada Research Chairs Program.

\section*{Author contributions}

A.W. formulated synthetic correlated diffusion imaging, and conceived and designed the study. V.S. and H.G. conducted the data preparation. H.G. performed the experiments and constructed data visualizations. M.A.H. was involved in reviewing the data and providing clinical interpretation. All authors performed the data analysis. All authors contributed to the writing and editing of the paper.

\section*{Additional information}
\textbf{Competing interests:} The authors declare no competing interests.

\pagebreak
\vspace{-0.1in}
\section*{Supplementary Information}

\captionsetup[table]{name=Supplementary Table}
\setcounter{table}{0}

\begin{table}[h]
    \vspace{-0.1in}
    \centering
    \caption{Statistical analysis of differences in area under the ROC curve between all modalities. ROC curve titles correspond to those in Fig.~2.}
    \small
    \label{tab:signif}
    \begin{tabular}{|c|c|c|c|c|c|c|c|}
        \hline
        \textbf{ROC Curve} & \textbf{Modalities} & $\mathbf{A_1}$ & $\mathbf{A_2}$ & $\mathbf{r_P}$ & $\mathbf{r_N}$ & $\mathbf{z}$ & $\mathbf{p}$ \textbf{(two-tailed)}\\\hline
        \parbox[t]{2mm}{\multirow{10}{*}{\rotatebox[origin=c]{90}{PCa vs. Healthy}}}
        & CDI$^s$, CDI$^s_t$ & 0.9167 & 0.9189 & 0.9982 & 0.9802 & 2.3137 & 0.0207 \\ \cline{2-8}
        & CDI$^s$, ADC & 0.9167 & 0.8859 & 0.1886 & 0.2850 & 13.0366 & 0.0000 \\ \cline{2-8}
        & CDI$^s$, T2w & 0.9167 & 0.7130 & 0.1406 & 0.0854 & 113.9934 & 0.0000 \\ \cline{2-8}
        & CDI$^s$, $k^{trans}$ & 0.9167 & 0.7461 & 0.0057 & 0.0339 & 66.0474 & 0.0000 \\ \cline{2-8}
        & CDI$^s_t$, ADC & 0.9189 & 0.8859 & 0.1994 & 0.2857 & 14.0423 & 0.0000 \\ \cline{2-8}
        & CDI$^s_t$, T2w & 0.9189 & 0.7130 & 0.1464 & 0.0861 & 116.4832 & 0.0000 \\ \cline{2-8}
        & CDI$^s_t$, $k^{trans}$ & 0.9189 & 0.7461 & 0.0041 & 0.0466 & 67.2891 & 0.0000 \\ \cline{2-8}
        & ADC, T2w & 0.8859 & 0.7130 & 0.4009 & 0.6696 & 100.0295 & 0.0000 \\ \cline{2-8}
        & ADC, $k^{trans}$ & 0.8859 & 0.7461 & 0.1428 & 0.0295 & 52.2421 & 0.0000 \\ \cline{2-8}
        & T2w, $k^{trans}$ & 0.7130 & 0.7461 & 0.1922 & 0.0527 & 16.4462 & 0.0000 \\ \hline\hline

        \parbox[t]{2mm}{\multirow{10}{*}{\rotatebox[origin=c]{90}{PCa vs. Ins-PCa}}}
        & CDI$^s$, CDI$^s_t$ & 0.7552 & 0.7555 & 0.9982 & 0.9823 & 0.1956 & 0.8449 \\ \cline{2-8}
        & CDI$^s$, ADC & 0.7552 & 0.6940 & 0.1886 & 0.1887 & 16.3557 & 0.0000 \\ \cline{2-8}
        & CDI$^s$, T2w & 0.7552 & 0.5535 & 0.1406 & 0.0037 & 70.1154 & 0.0000 \\ \cline{2-8}
        & CDI$^s$, $k^{trans}$ & 0.7552 & 0.6208 & 0.0057 & 0.1245 & 37.3343 & 0.0000 \\ \cline{2-8}
        & CDI$^s_t$, ADC & 0.7555 & 0.6940 & 0.1994 & 0.1940 & 16.5100 & 0.0000 \\ \cline{2-8}
        & CDI$^s_t$, T2w & 0.7555 & 0.5535 & 0.1464 & 0.0050 & 70.3044 & 0.0000 \\ \cline{2-8}
        & CDI$^s_t$, $k^{trans}$ & 0.7555 & 0.6208 & 0.0041 & 0.1134 & 37.3052 & 0.0000 \\ \cline{2-8}
        & ADC, T2w & 0.6940 & 0.5535 & 0.4009 & 0.6527 & 52.4005 & 0.0000 \\ \cline{2-8}
        & ADC, $k^{trans}$ & 0.6940 & 0.6208 & 0.1428 & 0.1707 & 20.4215 & 0.0000 \\ \cline{2-8}
        & T2w, $k^{trans}$ & 0.5535 & 0.6208 & 0.1922 & 0.0028 & 27.0110 & 0.0000 \\ \hline\hline

        \parbox[t]{2mm}{\multirow{10}{*}{\rotatebox[origin=c]{90}{PCa vs. Ins-PCa and Healthy}}}
        & CDI$^s$, CDI$^s_t$ & 0.9132 & 0.9153 & 0.9982 & 0.9825 & 2.2405 & 0.0251 \\ \cline{2-8}
        & CDI$^s$, ADC & 0.9132 & 0.8818 & 0.1886 & 0.1881 & 12.8230 & 0.0000 \\ \cline{2-8}
        & CDI$^s$, T2w & 0.9132 & 0.7100 & 0.1406 & 0.0238 & 110.9176 & 0.0000 \\ \cline{2-8}
        & CDI$^s$, $k^{trans}$ & 0.9132 & 0.7429 & 0.0057 & 0.1496 & 66.6417 & 0.0000 \\ \cline{2-8}
        & CDI$^s_t$, ADC & 0.9153 & 0.8818 & 0.1994 & 0.1966 & 13.8127 & 0.0000 \\ \cline{2-8}
        & CDI$^s_t$, T2w & 0.9153 & 0.7100 & 0.1464 & 0.0249 & 113.1965 & 0.0000 \\ \cline{2-8}
        & CDI$^s_t$, $k^{trans}$ & 0.9153 & 0.7429 & 0.0041 & 0.1420 & 67.6887 & 0.0000 \\ \cline{2-8}
        & ADC, T2w & 0.8818 & 0.7100 & 0.4009 & 0.6335 & 97.1469 & 0.0000 \\ \cline{2-8}
        & ADC, $k^{trans}$ & 0.8818 & 0.7429 & 0.1428 & 0.3370 & 55.6735 & 0.0000 \\ \cline{2-8}
        & T2w, $k^{trans}$ & 0.7100 & 0.7429 & 0.1922 & 0.0943 & 16.4136 & 0.0000 \\ \hline\hline

        \parbox[t]{2mm}{\multirow{10}{*}{\rotatebox[origin=c]{90}{PCa and Ins-PCa vs. Healthy}}}
        & CDI$^s$, CDI$^s_t$ & 0.8282 & 0.8345 & 0.9959 & 0.9802 & 9.3085 & 0.0000 \\ \cline{2-8}
        & CDI$^s$, ADC & 0.8282 & 0.8131 & 0.1550 & 0.2850 & 8.8399 & 0.0000 \\ \cline{2-8}
        & CDI$^s$, T2w & 0.8282 & 0.6810 & 0.0673 & 0.0854 & 108.2148 & 0.0000 \\ \cline{2-8}
        & CDI$^s$, $k^{trans}$ & 0.8282 & 0.6772 & 0.0442 & 0.0339 & 85.7535 & 0.0000 \\ \cline{2-8}
        & CDI$^s_t$, ADC & 0.8345 & 0.8131 & 0.1600 & 0.2857 & 12.5640 & 0.0000 \\ \cline{2-8}
        & CDI$^s_t$, T2w & 0.8345 & 0.6810 & 0.0728 & 0.0861 & 114.2608 & 0.0000 \\ \cline{2-8}
        & CDI$^s_t$, $k^{trans}$ & 0.8345 & 0.6772 & 0.0434 & 0.0466 & 90.1805 & 0.0000 \\ \cline{2-8}
        & ADC, T2w & 0.8131 & 0.6810 & 0.6176 & 0.6696 & 114.0235 & 0.0000 \\ \cline{2-8}
        & ADC, $k^{trans}$ & 0.8131 & 0.6772 & 0.2279 & 0.0295 & 79.4537 & 0.0000 \\ \cline{2-8}
        & T2w, $k^{trans}$ & 0.6810 & 0.6772 & 0.0054 & 0.0527 & 3.0826 & 0.0021 \\ \hline
    \end{tabular}
    \vspace{-1.1in}
\end{table}


\begin{thebibliography}{99}
\bibliographystyle{nature}

\bibitem{globocan21}
{H. Sung et al.}: \textbf{Global Cancer Statistics 2020: GLOBOCAN Estimates of Incidence and Mortality Worldwide for 36 Cancers in 185 Countries}. \emph{CA: A Cancer Journal for Clinicians} 2021, \textbf{71}(3), 209--249.

\bibitem{survival}
{Canadian Cancer Society}: \textbf{Canadian Cancer Statistics} 2020.

\bibitem{meta1}
{Ren et al}: \textbf{MRI of Prostate Cancer Antigen Expression for Diagnosis
  and lmmunotherapy}. \emph{PLoS ONE} 2012, \textbf{7}(60):e38350.

\bibitem{meta2}
{Jemal et al.}: \textbf{Cancer statistics}. \emph{CA Cancer J Clin} 2010,
  \textbf{60}:277--300.

\bibitem{meta3}
{Damber et al.}: \textbf{Prostate cancer}. \emph{Lancet} 2008,
  \textbf{371}:1710--1721.

\bibitem{meta4}
{Jani and Hellman}: \textbf{Early prostate cancer: clinical decision-making}.
  \emph{Lancet} 2003, \textbf{361}:1045--1053.

\bibitem{meta5}
Gronberg: \textbf{Prostate cancer epidemiology}. \emph{Lancet} 2003,
  \textbf{361}:859--864.

\bibitem{Stenman}
{Stenman et al.}: \textbf{Prostate-specific antigen}. \emph{Seminars in Cancer
  Biology} 1999, \textbf{9}:83--93.

\bibitem{Chou}
{Chou et al.}: \textbf{Screening for Prostate Cancer - A Review of the Evidence
  for the U.S. Preventive Services Task Force}. \emph{Annals of Internal
  Medicine} 2011.

\bibitem{Loeb}
{S. Loeb et al.}: \textbf{Overdiagnosis and overtreatment of prostate cancer}. \emph{Eur Urol.} 2014, \textbf{65}(6):1046-1055.

\bibitem{Norberg}
{Norberg et al.}: \textbf{The sextant protocol for ultrasound-guided core
  biopsies of the prostate underestimates the presence of cancer}.
  \emph{Urology} 1997, \textbf{50}:562--566.

\bibitem{Beerlage}
{Beerlage et al.}: \textbf{Correlation of transrectal ultrasound, computer
  analysis of transrectal ultrasound and histopathology of radical
  prostatectomy specimen}. \emph{Prostate Cancer and Prostatic Diseases} 2001,
  \textbf{4}:56--62.

\bibitem{PET1}
{Olafsen et al.}: \textbf{Targeting, imaging, and therapy using a humanized
  antiprostate stem cell antigen (PSCA) antibody}. \emph{J Immunother} 2007,
  \textbf{30}:396--405.

\bibitem{PET2}
{Lapi et al.}: \textbf{Assessment of an 18F-labeled phosphoramidate
  peptidomimetic as a new prostate-specific membrane antigen-targeted imaging
  agent for prostate cancer}. \emph{Nucl Med} 2009, \textbf{50}:2042--2048.

\bibitem{PET3}
{Lepin et al.}: \textbf{An affinity matured minibody for PET imaging of prostate
  stem cell antigen (PSCA)-expressing tumors}. \emph{Eur J Nucl Med Mol
  Imaging} 2010, \textbf{37}:1529--1538.

\bibitem{PET4}
{Giovacchini et al}: \textbf{PSA doubling time for prediction of 11C choline
  PET/CT findings in prostate cancer patients with biochemical failure after
  radical prostatectomy}. \emph{Eur J Nucl Med Mol Imaging} 2010,
  \textbf{37}:1106--1116.

\bibitem{PET5}
{Turkbey et al.}: \textbf{Imaging techniques for prostate cancer: implications
  for focal therapy}. \emph{Nat Rev Urol} 2009, \textbf{6}:191--203.

\bibitem{PIRADS}
{Weinreb et al.}: \textbf{PI-RADS Prostate Imaging-Reporting and Data System:2015,Version 2}. \emph{Eur Urol} 2016, \textbf{69} 16--40.

\bibitem{T2a}
{Khoo et al}: \textbf{Comparison of MRI with CT for the radiotherapy planning
  of prostate cancer: a feasibility study}. \emph{Br J Radiol.} 1999,
  \textbf{72}:590--597.

\bibitem{T2b}
{Debois et al}: \textbf{The contribution of magnetic resonance imaging to the
  three-dimensional treatment planning of localized prostate cancer}. \emph{Int
  J Radiat Oncol Biol Phys.} 1999, \textbf{45}:857--865.

\bibitem{T2c}
{Jackson et al}: \textbf{Distortion-corrected T2 weighted MRI: a novel approach
  to prostate radiotherapy planning}. \emph{The British Journal of Radiology}
  2007, \textbf{80}:926--933.

\bibitem{fMRI}
{Choi et al}: \textbf{Functional MR Imaging of Prostate Cancer}.
  \emph{RadioGraphics} 2007, \textbf{27}:63--75.

\bibitem{Koh}
{Koh and Padhani}: \textbf{Diffusion-weighted MRI: a new functional clinical
  technique for tumour imaging}. \emph{British Journal of Radiology} 2006,
  \textbf{79}:633--635.

\bibitem{Hosseinzadeh}
{Hosseinzadeh and Schwarz}: \textbf{Endorectal diffusion-weighted imaging in
  prostate cancer to differentiate malignant and benign peripheral zone
  tissue}. \emph{Journal of Magnetic Resonance Imaging} 2004,
  \textbf{20}:654--661.

\bibitem{Woodfield}
{Woodfield et al}: \textbf{Diffusion-Weighted MRI of Peripheral Zone Prostate
  Cancer: Comparison of Tumor Apparent Diffusion Coefficient With Gleason Score
  and Percentage of Tumor on Core Biopsy}. \emph{American Journal of
  Roentgenology} 2010, \textbf{194}:316--322.

\bibitem{Turkbey}
{Turkbey et al}: \textbf{Is Apparent Diffusion Coefficient Associated with
  Clinical Risk Scores for Prostate Cancers that Are Visible on 3-T MR Images?}
  \emph{Radiology} 2011, \textbf{258}:488--495.

\bibitem{Haider}
{Haider et al}: \textbf{Combined T2-weighted and diffusion-weighted MRI for
  localization of prostate cancer}. \emph{American Journal of Roentgenology}
  2007, \textbf{189}:323--328.

\bibitem{T2d}
{Barentsz et al}: \textbf{ESUR prostate MR guidelines 2012}. \emph{Eur Radiol.}
  2012, \textbf{22}(4):746--757.

\bibitem{Quon}
{Quon et al.}: \textbf{False positive and false negative diagnoses of prostate cancer
at multi-parametric prostate MRI in active surveillance}. \emph{Insights Into Imaging} 2015, \textbf{6}:449--463.

\bibitem{DCE}
{Berman et al.}: \textbf{DCE MRI of prostate cancer}. \emph{Abdominal radiology} 2016, \textbf{41}(5):844--853.

\bibitem{CDI}
{Wong et al.}:
  \textbf{Correlated Diffusion Imaging}. \emph{BMC Medical Imaging} 2013, \textbf{13}, 1--7.

\bibitem{Khalvati15}
{Khalvati, Wong, and Haider}: \textbf{Automated prostate cancer detection via comprehensive multi-parametric magnetic resonance imaging texture feature models}. \emph{BMC Medical Imaging} 2015, 15--27.

\bibitem{Litjens14}
{Litjens et al.}: \textbf{Computer-Aided Detection of Prostate Cancer in MRI}. \emph{IEEE Transactions on Medical Imaging} 2014, \textbf{33}(5):1083--1092.

\bibitem{Cuocolo21}
{Cuocolo et al.}: \textbf{Quality control and whole-gland, zonal and lesion annotations for the PROSTATEx challenge public dataset.} \emph{Eur. J. Radiol.} 2021, 109647.

\bibitem{Ploussard11}
{Ploussard et al.}: \textbf{The contemporary concept of significant versus insignificant prostate cancer}. \emph{Eur Urol.} 2011, \textbf{60}(2):291–303.

\bibitem{Stejskal}
{Stejskal and Tanner}: \textbf{Spin Diffusion Measurements: Spin Echoes in the
  Presence of a Time-Dependent Field Gradient}. \emph{The Journal of Chemical
  Physics} 1965, \textbf{42}:288.

\bibitem{Bihan}
{Le Bihan and Breton}: \textbf{Imagerie de diffusion in-vivo par resonance
  magnetique nucleaire}. \emph{C R Acad Sci} 1985, \textbf{301}:1109--1112.

\bibitem{Hanley}
{Hanley and McNeil}: \textbf{A Method of Comparing the Areas under Receiver Operating Characteristic Curves Derived from the Same Cases}. \emph{Radiology} 1983, \textbf{148}(3):839-843.

\end{thebibliography}
\end{document}